\documentclass[a4paper,12pt]{article}

\usepackage[hmargin=.7in,vmargin=1.1in]{geometry}
\usepackage{indentfirst}
\usepackage{amsfonts}
\usepackage{mathrsfs}
\usepackage{amsmath}
\usepackage{amssymb}
\usepackage{cite}
\usepackage{xcolor}
\usepackage{mathtools}
\usepackage{tensor}
\usepackage{graphicx}% Include figure files
\usepackage{bm}% bold math
\usepackage{upgreek}
\usepackage{braket}
\usepackage{color,soul}
\usepackage{csquotes}
\usepackage{caption}
\usepackage{subcaption}
\usepackage{tabularx}
\usepackage[section]{placeins}
\usepackage{enumerate}
\usepackage{comment}

\usepackage{makecell}  % 单元格内换行必备
\usepackage{multirow}  % 纵向合并单元格
\usepackage{array}     % 增强表格功能

\usepackage{diagbox}
\newcommand{\longslash}{\diagbox[width=5.5em, height=3.2em]{}{}}
\newcommand{\rd}{\mathrm{d}}

\newcommand{\pfrac}[2]{\left(\frac{\partial #1}{\partial #2}\right)}
\newcommand{\ppfrac}[3]{\left(\frac{\partial^2 #1}{\partial #2 \partial #3}\right)}

\newcommand{\blue}[1]{\textcolor{blue}{#1}}
\renewcommand{\blue}[1]{\textcolor{black}{#1}}

\usepackage[bookmarksnumbered=true,bookmarksopen=true]{hyperref}
 \hypersetup{colorlinks,
             linkcolor=[rgb]{0,0.3,0.6}, 
             citecolor=[rgb]{0,0.3,0.6}, 
             urlcolor=[rgb]{0,0.3,0.6}}

\allowdisplaybreaks

\title{\Large Large black holes at a finite distance:\\
Quasi-local restricted phase space formalism}

\author{Bai-Hao Huang\thanks{huangbaihao@mail.nankai.edu.cn} 
~and 
Liu Zhao\thanks{Corresponding author, lzhao@nankai.edu.cn}\\
{\em School of Physics, Nankai University, 
Tianjin 300071, China}}

\numberwithin{equation}{section}

\date{}

\begin{document}

\maketitle

\begin{abstract}

We extend the restricted phase space formalism for spherically symmetric black hole 
solutions of Einstein-Maxwell theory 
to the quasi-local regime, with the static observers located at a finite radial 
distance. The first law and Euler relation for the RN and RN-AdS black holes 
are proved to hold, but only with the inclusion of an extra pair of 
thermodynamic variables, {\em i.e.} the pressure $P$ and the area $A$ of the 
codimension-2 hypersurface on which the observers reside. 
For the RN black holes, the quasi-local behavior is analyzed in detail. 
It turns out that the RN black holes in the quasi-local description behaves 
significantly different from itself in the asymptotic description, 
but is extremely similar to the RN-AdS black holes in the asymptotic description, 
{\em e.g.} allowing for isocharge temperature-entropy phase transitions 
and lack of isovoltage temperature-entropy phase transitions. In the neutral limit, 
the Hawking-Page-like transitions appear in the quasi-local description which 
is absent in the asymptotic description. 
\blue{The quasi-local behavior of RN-AdS black holes is also briefly discussed, 
which is qualitatively identical to the asymptotic description with only 
quantitative differences.}

\vspace{1em}

\end{abstract}

%\tableofcontents

%%%%%%%%%%%%%%%%%%%%%%%%%%%%%%%
\section{Introduction}
\label{sec:intr}
%%%%%%%%%%%%%%%%%%%%%%%%%%%%%%

Black holes (BHs) are important objects predicted by general relativity.
Since the pioneering works on black hole thermodynamics (BHT)
\cite{Bekenstein:1972tm,Bekenstein:1973ur},
the thermodynamic properties of BHs have been extensively investigated 
from various perspectives, such as the four basic laws of BHT \cite{Bardeen:1973gs}, 
the role of gravitational path integral\cite{Gibbons:1976ue} in BHT,
and the phase structure analysis
\cite{Hawking:1982dh,Chamblin:1999tk,Chamblin:1999hg}, {\em etc}.
Though the four laws of BHT are very similar to those of ordinary matter,
the Euler homogeneity that is crucial in extensive thermodynamics is
absent except in some subtle exceptional cases 
\cite{Tian1,Tian2}. The restricted phase space (RPS) formalism
\cite{Zeyuan:2021uol} is a relatively new formalism for BHT with 
complete Euler homogeneity, which allows for a proper definitions of 
some standard concepts in ordinary thermodynamics, {\em e.g.} subsystem, 
equilibrium conditions and stability criteria {\em etc} in BHT \cite{Wang:2022err}. 
Subsequent developments
\cite{Gao:2021xtt,Wang:2021cmz,Wang:2022err,Bai:2023wjm,Ali:2023ppg,
Ladghami:2023ccf,Sadeghi:2022jlz,Tripathy:2024rgt,Wang:2022bqu,Jian:2024wjb}
seem to indicate that this new formalism is universally applicable to 
all black hole solutions in various different gravitational theories. 
In the RPS formalism, inspired by AdS/CFT duality \cite{Maldacena:1997re},
the chemical potential related to the on-shell Euclidean action $I_E$ 
via the relation $\mu=T_{\mathrm{H}}I_E/N$ and the effective number of 
microscopic degrees of freedom $N=L^{d-2}/G$ are introduced as 
a novel conjugate pair of thermodynamic variables,
where $T_{\mathrm{H}}$ is the Hawking temperature,
$d$ is the spacetime dimension, and $L$ is a constant characteristic length scale.
The cosmological constant $\Lambda$ is kept unchanged in this formalism,
while the Newton constant $G$ is allowed to vary in order that $N$ becomes a variable.

Observers play an outstanding role in both general relativity and thermodynamics, 
because different observers can observe different phenomena while observing the same 
physical process. In the previous studies on BHT, the observers are rarely explicitly 
mentioned, but the use of ADM energy and Hawking temperature in 
such studies implies that the observers are implicitly placed at the asymptotic 
infinity of the spacetime. However, there are important reasons to reconsider BHT 
from the view point of observers placed at a finite distance away from the black hole. 
This is particularly important if we consider the potential observationary verification 
of BHT, because any observation will most probably be carried out in our solar 
system, with the black hole not too far away comparing to the observable radius 
of the universe. Historically, the study of black hole 
observables with respect to observers at a finite distance is known as the quasi-local
description \cite{Brown:1992br,York:1986it}, \blue{however in the absence of 
Euler homogeneity and thermodynamic extensivity}. The aim of the present work 
is to extend the RPS formalism into the regime of quasi-local description, 
while still maintaining the full merits of the RPS formalism itself, especially 
the full Euler homogeneity and the universal applicability. 

There are many kinds of definitions of quasi-local energy 
--- the total energy possessed by the gravitational field
within the spacetime region enclosed by a finite closed hypersurface
--- such as the Komar integral at a finite closed hypersurface,
Kijowski-Liu-Yau energy, Epp energy and Misner-Sharp energy \cite{Szabados:2009eka}.
Among them, the first law and the Smarr relation of Schwarzschild BHs for the 
static observers at the finite radius $r=R$ are studied in the Brown-York 
quasi-local description \cite{Brown:1992br,York:1986it}, and expressed as
\begin{equation}
\mathrm{d}E=T_{R}\mathrm{d}S-P\mathrm{d}A,
\quad
E=2\left(T_RS-PA\right),
\label{eq:yorks}
\end{equation}
where $T_{R}=T_{\mathrm{H}}/\sqrt{1-2GM/R}$ is the Tolman temperature 
measured by such static observers, $P$ is the surface pressure
with the surface area $A=4\pi R^2$ being its thermodynamic conjugate.
These results have also been extended to other BHs \cite{Brown:1994gs,Braden:1990hw,
Carlip:2003ne,Akbar:2004ke,Lundgren:2006kt,Lee:2018hrd,Villalba:2019sxl,
Fernandes:2025vzb}.
Besides, the quasi-local energy density of Schwarzschild BHs
for different finite-distance observers are studied in \cite{Booth:1998eh},
and the Lorentz boost relation between the quasi-local energy density
and momentum density is studied in \cite{Brown:2000dz}.
Recently, the Brown-York quasi-local description
is combined with the extended phase space formalism
\cite{Simovic:2018tdy,Fontana:2018drk,Wang:2020hjw,Wang:2021llu,Zhao:2020nrx}.
As can be seen in Eq.~\eqref{eq:yorks} and its counterparts in 
\cite{Simovic:2018tdy,Fontana:2018drk,Wang:2020hjw,Wang:2021llu,Zhao:2020nrx},
Euler homogeneity is still missing in the Brown-York quasi-local description
and its combination with the extended phase space formalism. 
\blue{Since the Euler homogeneity is missing in all previous studies 
about the quasi-local description, a key feature of phase transition ({\em i.e.} 
scaling properties) cannot be recovered in such studies. Therefore, the 
description about phase transitions in such works is not sufficiently convincing 
and needs to be reconsidered in the presence of Euler homogeneity.}
By extending the RPS formalism into the regime of quasi-local 
description, we will achieve a quasi-local RPS formalism of BHT with full
Euler homogeneity. For comparison, the original RPS formalism 
may be rephrased as asymptotic RPS formalism.  We will see that the 
BHs can have different thermodynamic behaviors 
in the quasi-local and asymptotic RPS formalisms. 
For illustration purposes, we will consider only spherically symmetric solutions 
in the Einstein-Maxwell theory with or without a negative cosmological constant, 
{\em i.e.} the RN or RN-AdS black holes in order to maintain simplicity 
of the presentation.

%%%%%%%%%%%%%%%%%%%%%%%%%%%%%%%
\section{Euclidean action and quasi-local observables}
\label{sec:quasi}
%%%%%%%%%%%%%%%%%%%%%%%%%%%%%%

The on-shell Euclidean action plays a prominent role in the RPS formalism. 
In the quasi-local description, the action must be described as an integration over 
a finite region $\mathcal{M}$ in the spacetime $M$ with a closed boundary hypersurface 
$\partial \mathcal{M}$ consisting of three parts. In Schwarzschild-like coordinates $x^\mu=(t,r,\theta,\phi)$, 
these three parts can be specified respectively as
\blue{
\begin{align}
\Sigma_{t_i} &= \{ x^\mu\in M| t=t_i, r\leq R\},\\
\Sigma_{t_f} &= \{ x^\mu\in M| t=t_f, r\leq R\},\\
B &= \{x^\mu \in M | t_i \leq t\leq t_f, r=R\}.
\end{align}
}
Clearly, the intersection $\Omega_t$ of any constant-time hypersurface with $B$ 
is a codimension-2 spherical surface of radius $r=R$. Thus $\Omega_{t_i}=\Sigma_{t_i} 
\cap B$ and $\Omega_{t_f}=\Sigma_{t_f} \cap B$ are both codimension-2 spherical surfaces.

In order that the variational process works consistently, 
the action needs to be consisted of a bulk and a boundary term. In the present setting, 
the boundary contribution comes only from $B$, {\em i.e.}
\begin{equation}
I_{\mathcal M} = \frac{1}{16 \pi G} \int_{\mathcal{M}} \rd^4 x   
    \sqrt{- g}\left({\mathcal R}-2\Lambda-F_{\mu\nu}F^{\mu\nu}\right) 
	+ \frac{1}{8 \pi G} \int_{B} 
    \rd^3 x \sqrt{- \gamma} \Theta,
\label{eq:action}
\end{equation}
where $g,~\gamma$ respectively are the determinant of the metrics on 
$\mathcal{M}$ and $B$ respectively, ${\mathcal R}$ is the Ricci curvature scalar,
$\Lambda$ is the cosmological constant (which is either negative or zero in the 
subsequent considerations), $F_{\mu\nu}=\nabla_{\mu}A_{\nu}-\nabla_{\nu}A_{\mu}$
is the Maxwell field strength, and 
$\Theta=\gamma^{\alpha\beta}\Theta_{\alpha\beta}$ is the trace of the extrinsic 
curvature $\Theta_{\alpha\beta}$ of $B$. Please be reminded that, 
in writing the action \eqref{eq:action}, the electromagnetic potential $A_\mu$ 
has been properly rescaled, so that the $\mathcal R$ and $F^2$ terms have a common 
overall factor $1/16\pi G$. Without loss of generality, the spherical symmetric solution 
to the above theory can be cast into a unified form,
\begin{align}
\rd s^2=-f(r)\rd t^2 +f^{-1}(r)\rd r^2 +r^2(\rd \theta^2+\sin^2\theta\rd\phi^2),
\label{lelem1}
\end{align}
where $f(r)$ can take different form for different black hole solutions.

After Euclideanization, the Euclidean time $t_E={\rm i}\, t$ needs to have a finite period 
$\beta_{\rm H}=1/T_{\rm H}$ in order to resolve the conical singularity at 
the event horizon. It is natural to take $t_{Ei} =0, 
t_{Ef} = \beta_{\rm H}$. For $\Lambda \leq 0$, directly substituting the 
Euclideanized metric $g^{(E)}_{\mu\nu}$ into the action $I_{\mathcal M}$ 
would normally lead to a divergent result. Therefore, we need to introduce a 
background counter term in the action to get a finite result 
after Euclideanization. The total action reads
\begin{align}
I = I_{\mathcal M} - I_{{\mathcal M}_{0}},
\label{eq:eaction}
\end{align}
where ${\mathcal M}_{0}$ is a finite region in the background spacetime $M_0$ 
with the same boundary geometry as $\mathcal M$, 
and $I_{{\mathcal M}_{0}}$ is $I_{\mathcal M}$ with $\mathcal M$ replaced by 
${\mathcal M}_{0}$. The background spacetime $M_0$ 
is taken to be the pure AdS spacetime with the same cosmological constant 
for asymptotically AdS or the Minkowski spacetime for asymptotically flat cases 
\cite{Gibbons:1976ue,Braden:1990hw,Chamblin:1999tk,Brown:1992br,
Balasubramanian:1999re}. In either cases, the line element on $M_0$ can be 
written in a unified form
\begin{align}
\rd s_0^2=-f_0(r)\rd \tau^2 +f_0^{-1}(r)\rd r^2 +r^2(\rd \theta^2+\sin^2\theta\rd\phi^2),
\label{lelem2}
\end{align}
where $\tau$ is the time coordinate on $M_0$. Evidently, 
$\displaystyle f_0(r) =1+\frac{r^2}{\ell^2}$ (where $\ell^2 =-3/\Lambda$)
for the pure AdS case and $f_0(r)=1$ for the Minkowski case.

According to the line elements \eqref{lelem1} and \eqref{lelem2}, 
the induced line elements on $B$ and $B_0$ can be written respectively as
\begin{align*}
&\rd s_{B}^2 = - f(R) \rd t^2 + R^2(\rd \theta^2+\sin^2\theta\rd\phi^2),\\
&\rd s_{B_0}^2 = -f_0(R) \rd \tau^2 + R^2(\rd \theta^2+\sin^2\theta\rd\phi^2).
\end{align*} 
In order to match the geometry of $B$ and $B_0$, we need to require
$f_0(R) \rd \tau^2= f(R) \rd t^2$. This in turn implies that
the period of the Euclidean time $\tau_E ={\rm i}\,\tau$ must be taken to be
$\beta_0 = \beta_{\rm H}\frac{\sqrt{f(R)}}{\sqrt{f_0(R)}}$.

Since $B$ and $B_0$ has the same metric, the boundary terms in both 
$I_{\mathcal M}$ and $I_{\mathcal M_0}$ have contributions to the boundary stress
tensor $\tau^{\alpha\beta}$ \cite{Brown:1992br}:
\begin{equation}
\tau^{\alpha\beta} =\frac{2}{\sqrt{- \gamma}}
\frac{\delta I}{\delta \gamma_{\alpha\beta}}
=\frac{2}{\sqrt{- \gamma}} 
\left(\pi^{\alpha\beta}_{\mathcal M} - \pi^{\alpha\beta}_{\mathcal M_0}\right),
\end{equation}
where
\begin{equation}
\pi^{\alpha\beta}_{\mathcal M}=-\frac{\sqrt{-\gamma}}{16\pi G} \left(\Theta^{\alpha\beta} 
- \Theta \gamma^{\alpha\beta}\right),
\end{equation}
$\pi^{\alpha\beta}_{\mathcal M_0}$ is its counterpart on $\mathcal M_0$, 
and each one of the the indices $\alpha, \beta$ can take the values $0, 2, 3$ 
corresponding to the coordinates $t,\theta,\phi$.

From the boundary stress tensor $\tau^{\alpha\beta}$, 
the quasi-local energy density $\varepsilon$ and the spatial stress tensor $s^{ab}$ 
can be extracted, {\em i.e.}
\begin{equation}
\varepsilon  = \tau_{\alpha\beta} u^{\alpha}u^{\beta},
\quad s^{ab}  =  \sigma^{a\alpha} \sigma^{b\beta}  \tau_{\alpha\beta},
\end{equation}
where $\sigma^{\alpha\beta}$ is the induced metric on $\Omega_t$, 
$a,b$ can only take the values $2,3$ corresponding to $\theta,\phi$
respectively, and $u^{\alpha}$ is $u^\mu$ --- the four-velocity of the static observers located 
at $r=R$ --- projected onto the submanifold $B$, {\em i.e.}
$u^{\alpha}=\left(1/\sqrt{f\left(R\right)},0,0\right)$. 
By integration over the hypersurface $\Omega_t$,
we obtain the quasi-local energy $E$,
\begin{equation}
E=\int_{\Omega_t}\mathrm{d}^2x\sqrt{\sigma}\varepsilon.
\end{equation}
The surface pressure $P$ is defined by
\begin{equation}
P=\frac{1}{2}\sigma_{ab}s^{ab}.
\end{equation}
Inserting the relevant metrics into the above definitions, 
the quasi-local energy $E$ and surface pressure $P$
can be explicitly evaluated \cite{Booth:1998eh}, yielding 
\blue{
\begin{align}
E=\frac{R}{G}\left(\sqrt{f_0\left(R\right)}-\sqrt{f\left(R\right)}\right),
\label{eq:quasi-E}
\end{align}
\begin{align}
P=\frac{1}{8\pi G}\left( 
\blue{-}\frac{\sqrt{f_0\left( R \right)}
-\sqrt{f\left( R \right)}}{R}
+\frac{f^{\prime}\left( R \right)}
{2\sqrt{f\left( R \right)}}
-\frac{f_{0}^{\prime}\left( R \right)}
{2\sqrt{f_0\left( R \right)}} \right) ,
\label{eq:quasi-P}
\end{align}
}
where primes denote the derivatives with respects to the arguments. 
In the quasi-local description of BHT, the quasi-local energy $E$ will be interpreted 
as the internal energy. Meanwhile, the surface pressure $P$ 
is considered to be a thermodynamic quantity, 
with the area $A=4\pi R^2$ of the codimension-2 hypersurface $\Omega_t$
being its conjugate quantity. Later on, we will see that 
$P$ and $A$ need to be rescaled as $\hat{P}=PG,~\hat{A}=A/G=4\pi R^2/G$ 
in the RPS formalism.

As is well known, the temperature measured by static observers at $r=R$
is the the Tolman temperature $T_R$, which differs from the 
Hawking temperature $T_{\mathrm{H}}$ by a redshift factor
$\blue{1/\sqrt{f(R)}}$
\cite{York:1986it}
\begin{equation}
T_R=\frac{T_{\mathrm{H}}}{\sqrt{f\left(R\right)}}.
\label{TR215}
\end{equation}
\begin{comment}
This result can be understood in two ways.
On the one hand,
after wick rotating $t=-\mathrm{i}t_E$,
the Euclidean time $t_E$ needs the period
$\beta_{\mathrm{H}}$,
and the proper length of the $S^1$ of the boundary
is $\beta _R=\int_0^{\beta _{\mathrm{H}}}{\sqrt{g_{t_Et_E}}\mathrm{d}t_E}
=\beta _{\mathrm{H}}\sqrt{f\left( R \right)}$
\cite{York:1986it}.
On the other hand,
the Hawking effect for the static observers
in the BHs spacetime can be related to
the Unruh effect in the 
higher dimensional flat spacetime
\cite{Deser:1998xb}.
\end{comment}
However, the thermodynamic conjugate to the temperature, {\em i.e.} the entropy
of the black hole, remains unchanged and is given by Bekenstein-Hawking entropy
$\displaystyle S = \frac{\pi r_h^2}{G}$ with $r_h$ being the 
radius of the event horizon.

Besides the two conjugate pairs $(T_{R}, S)$ and $(\hat P, \hat A)$, 
there is a third pair $(\hat Q, \hat\Phi)$ of conjugate variables
for charged BHs, where 
\begin{align}
&\hat{Q}=\frac{QL}{\sqrt{G}},\quad 
\hat{\Phi}_R=\frac{\sqrt{G}\Phi_R}{L},
\label{PhiRdef}
\end{align}
in which $Q$ is the electric charge related to the electromagnetic field 
$A_{\mu}$ via $A_{\mu}=\left(Q/r,0,0,0\right)$, and $\Phi_R$ 
is the electric potential measured by the static observers at $r=R$
\cite{Braden:1990hw}:
\begin{equation}
\Phi _R=-u^{\mu} A_{\mu} \big|_{r=r_h}^{r=R}
=\frac{Q}{\sqrt{f\left( R \right)}}\left( \frac{1}{r_h}-\frac{1}{R} \right).
\end{equation}

The last pair $(\mu_R, N)$ of conjugate thermodynamic variables is specific 
to the RPS formalism. In four spacetime dimensions, $N$ is given as 
$N=L^2/G$, whereas $\mu_R$ is linearly related to the Euclidean action $I_E$ 
but is no longer proportional to $I_E$ in the quasi-local description. 
\begin{comment}
\blue{According to Eq.~\eqref{eq:action}, the Euclidean action $I_E$ evaluates
the solutions with same Tolman temperature $T_R$, electric potential $\hat{\Phi}_R$
for the static observers fixed at $r=R$
at the saddle point approximation.
Therefore, the thermodynamic potential $\Psi_R$, related to Euclidean action $I_E$,
satisfies
\begin{equation}
\Psi_R(T_R,\hat{\Phi}_R,\hat{A},N)=T_RI_E=E-T_RS-\hat{\Phi}_R\hat{Q},
\end{equation}
The Gibbs energy $G_R$ is obtained from the Legendre transformation of $\Psi_R$,
\emph{i.e.} $G_R=\Psi_R+\hat{P}\hat{A}$,
while the chemical potential $\mu_R$ is defined as the Gibbs free energy per average particle.
}
\end{comment}
Concretely, we have 
\begin{align}
\mu_R=\frac{T_R I_E+\hat{P}\hat{A}}{N}, \label{muR218}
\end{align}
where the suffix $R$ indicates that the chemical potential $\mu_R$ is measured by 
static observers located at $r=R$. The 
\blue{origin of Eq.\eqref{muR218} will be explained in Subsection \ref{sec:RN}, 
and its explicit value} can only be calculated 
by use of the explicit metric functions $f(r)$ and $f_0(r)$. The corresponding 
process will be illustrated in the example cases of RN and RN-AdS BHs in the next 
section.

\blue{In order to have a more intuitive comparison about the asymptotic and quasi-local 
RPS formalisms, the definitions and scaling properties
of thermodynamic quantities in the asymptotic formalism and 
quasi-local RPS formalism are listed in
Table \ref{tab:thermo_comparison}.}

\begin{table}[htbp]
    \centering
    \setlength{\tabcolsep}{4pt} 
    \resizebox{\textwidth}{!}{
        \renewcommand{\arraystretch}{1.8} 
        \footnotesize 
        \begin{tabular}{|c|c|c|c|c|c|c|}
            \hline
             & \multicolumn{2}{c|}{\makecell[c]{\textbf{Asymptotic RPS} \\ \textbf{Formalism}}} & \multicolumn{2}{c|}{\makecell[c]{\textbf{Quasi-local RPS} \\ \textbf{Formalism}}} & \multicolumn{2}{c|}{\makecell[c]{\textbf{Conjugate} \\ \textbf{Quantity}}} \\ \cline{2-7} 
             & \textbf{Definition} & \textbf{Scaling} & \textbf{Definition} & \textbf{Scaling} & \textbf{Definition} & \textbf{Scaling} \\ \hline
            
            Energy & $M$ & $M \to \lambda M$ & $E: \mathrm{Eq}.\eqref{eq:quasi-E}$ & $E \to \lambda E$ & \longslash & \longslash \\ \hline
            
            Temperature & $T_{\mathrm{H}}$ & $T_{\mathrm{H}} \to T_{\mathrm{H}}$ & $T_R = \dfrac{T_{\mathrm{H}}}{\sqrt{f(R)}}$ & $T_R \to T_R$ & $S = \dfrac{\pi r_h^2}{G}$ & $S \to \lambda S$ \\ \hline
            
            \makecell[c]{Electric \\ Potential} & $\hat{\Phi} = \dfrac{\sqrt{G}Q}{L r_h}$ & $\hat{\Phi} \to \hat{\Phi}$ & $\hat{\Phi}_R: \mathrm{Eq}.\eqref{PhiRdef}$ & $\hat{\Phi}_R \to \hat{\Phi}_R$ & $\hat{Q} = \dfrac{QL}{\sqrt{G}}$ & $\hat{Q} \to \lambda \hat{Q}$ \\ \hline
            
            \makecell[c]{Surface \\ Pressure} & \longslash & \longslash & $\hat{P} = PG, \ P: \mathrm{Eq}.\eqref{eq:quasi-P}$ & $\hat{P} \to \hat{P}$ & $\hat{A} = \dfrac{4\pi R^2}{G}$ & $\hat{A} \to \lambda \hat{A}$ \\ \hline
            
            \makecell[c]{Chemical \\ Potential} & $\mu = \dfrac{T_{\mathrm{H}} I_E}{N}$ & $\mu \to \mu$ & $\mu_R = \dfrac{T_R I_E + \hat{P}\hat{A}}{N}$ & $\mu_R \to \mu_R$ & $N = \dfrac{L^2}{G}$ & $N \to \lambda N$ \\ \hline
            
            First Law & \multicolumn{2}{c|}{$\mathrm{d}M = T_{\mathrm{H}} \mathrm{d}S + \hat{\Phi} \mathrm{d}\hat{Q} + \mu \mathrm{d}N$} & \multicolumn{2}{c|}{$\mathrm{d}E = T_R \mathrm{d}S - \hat{P} \mathrm{d}\hat{A} + \hat{\Phi}_R \mathrm{d}\hat{Q} + \mu_R \mathrm{d}N$} & \longslash & \longslash \\ \hline
            
            Euler Relation & \multicolumn{2}{c|}{$M = T_{\mathrm{H}} S + \hat{\Phi} \hat{Q} + \mu N$} & \multicolumn{2}{c|}{$E = T_R S - \hat{P} \hat{A} + \hat{\Phi}_R \hat{Q} + \mu_R N$} & \longslash & \longslash \\ \hline
        \end{tabular}
    }
    \caption{Comparison of thermodynamic quantities in asymptotic and quasi-local RPS formalisms}
    \label{tab:thermo_comparison}
\end{table}

%%%%%%%%%%%%%%%%%%%%%%%%%%%%%%
\section{Quasi-local RPS formalism}
%%%%%%%%%%%%%%%%%%%%%%%%%%%%%%

In this section, we shall consider two simple example cases, {\em i.e.} 
the RN and RN-AdS BHs and present the explicit values of all quasi-local observables.
More importantly, we will show that the first law and the Euler homogeneity relation
hold simultaneously in the quasi-local RPS formalism.

\subsection{RN black hole}
\label{sec:RN}
%%%%%%%%%%%%%%%%%%%%%%%%%%%%%%

For the RN black hole, we have 
\begin{equation}
f\left(r\right)=1-\frac{2GM}{r}+\frac{GQ^2}{r^2},\quad
~f_0\left(r\right)=1.
\end{equation}
Using these metric functions, the quasi-local energy $E$, 
the rescaled surface pressure $\hat{P}$ and the corresponding surface area $\hat{A}$,
the Tolman temperature $T_R$ and the entropy $S$,
the rescaled electric potential $\hat{\Phi}_R$ and the electric charge $\hat{Q}$
are respectively evaluated to be 
\begin{align}
E&=\frac{R}{G}\left( 1-\sqrt{f(R)} \right) ,
\label{QE}
\\
\hat{P}&=\frac{1}{8\pi R}\left( \frac{1-GM/R}
{\sqrt{f\left( R \right)}}-1 \right) ,
\quad \hat{A}=\frac{4\pi R^2}{G},
\label{RN-P-A}
\\
T_R&=\frac{T_{\mathrm{H}}}
{\sqrt{f\left( R \right)}}
=\frac{r_{h}^{2}-GQ^2}{4\pi r_{h}^{3}
\sqrt{f\left( R \right)}},
\quad\,\, S=\frac{\pi r_{h}^{2}}{G},
\label{RN-T-S}
\\
\hat{\Phi}_R&=\frac{\sqrt{G}Q}
{L\sqrt{f\left( R \right)}}
\left( \frac{1}{r_h}-\frac{1}{R} \right) ,
\quad\quad\, \hat{Q}=\frac{QL}{\sqrt{G}},
\label{RN-Phi-Q}
\end{align}

The Euclidean action $I_E$ in this case is evaluated to be
\cite{Gibbons:1976ue,Braden:1990hw,Chamblin:1999tk,Brown:1992br}
\begin{align}
I_E
%&=-\frac{1}{16\pi G}\int_{\mathcal{M}} {\mathrm{d}^4x\sqrt{g} 
%\left(R-F_{\mu\nu}F^{\mu\nu}\right)}
%-\frac{1}{8\pi G}\int_B{\mathrm{d}^3x \sqrt{\gamma}\left( \Theta -\Theta _0 \right)}
%\nonumber
%\\
&=\frac{\beta _{\mathrm{H}}R}{G}
\left( \sqrt{f(R)} -1+\frac{3GM}{2R}-\frac{GQ^2}{2Rr_h} \right).
\label{RN-IE}
\end{align}
\blue{The exponential of $-I_E$ equals to the partition function of the gravitational 
path integral in the saddle point approximation, which in turn is identified with the 
partition function of the RN black hole regarded as a thermodynamical object. 
If we were studying ordinary matter systems, the latter partition function is usually 
identified with the exponential of the negative Helmholtz free energy divided by the 
temperature.  However, the RN black hole in the quasi-local RPS formalism is a 
thermodynamic system with more than three thermodynamic degrees of freedom, the ordinary 
terminologies for thermodynamic potentials become insufficient. Therefore, we prefer 
to reserve the term {\em Helmholtz free energy} for the quantity $F_R = E- T_R S$ 
and refer to the novel thermodynamic potential on the exponent of the latter 
partition function as $\Psi_R$.
Thus we have} the effective field theoretic relation 
\begin{align}
\exp\left(-\frac{\Psi_R}{T_R}\right) =\exp(-I_E).
\end{align}
\blue{It follows that} the thermodynamic potential $\Psi_R$ \blue{evaluates to be}
\begin{equation}
\Psi_R=T_RI_E=E-T_RS-\hat{\Phi}_R\hat{Q}.
\label{PsiR}
\end{equation}
\blue{Since the internal energy $E$ regarded as a thermodynamic potential 
takes $(S, \hat{A}, \hat{Q}, N)$ as its adapted independent variables, 
and Eq.\eqref{PsiR} viewed as a Legendre transformation indicates that the 
independent variables for $\Psi_R$ are $(T_R, \hat{A}, \hat{\Phi}_R, N)$. 
We need a further Legendre transformation to get the Gibbs free energy $G_R$, 
which takes $(T_R, \hat{P}, \hat{\Phi}_R, N)$ as independent variables, {\em i.e.}}
\begin{equation}
G_R=\Psi_R+\hat{P}\hat{A}
=\frac{R}{2G}\left[ 1-\sqrt{f(R)}
+\frac{GQ^2}{R\sqrt{f\left( R \right)}}
\left( \frac{1}{R}-\frac{1}{r_h} \right) \right].
\end{equation}
Finally, the quasi-local chemical potential $\mu_R$ is given as
\begin{equation}
\mu_R=\frac{G_R}{N}
=\frac{R}{2L^2}\left[ 1-\sqrt{f(R)}
+\frac{GQ^2}{R\sqrt{f\left( R \right)}}
\left( \frac{1}{R}-\frac{1}{r_h} \right) \right].
\end{equation}

Using the above data, it is now straightforward to prove the following relations:
\begin{align}
\rd E&=T_R\rd S-\hat{P}\rd \hat{A}+\hat{\Phi}_R\rd\hat{Q}+\mu_R\rd N,
\label{RPS1st}
\\
E&=T_RS-\hat{P}\hat{A}+\hat{\Phi}_R\hat{Q}+\mu_RN.
\label{RPSeuler}
\end{align}
Eq.\eqref{RPS1st} is the first law of BHT in the quasi-local RPS formalism, while 
Eq.\eqref{RPSeuler} is the Euler relation in the quasi-local RPS formalism. 
We see that the quasi-local RPS formalism contains an extra thermodynamic 
degree of freedom represented by the surface area $\hat A$. 
Let us remark that the inclusion of $\hat A$ together with its thermodynamic 
conjugate $\hat P$ in the first law is not surprising, because these 
variables were already considered as thermodynamic variables in the original 
quasi-local description \cite{Brown:1992br,York:1986it}. \blue{Please also be reminded that
even when the quantities $\hat A$ and $N$ are fixed via external constraints, 
these are still well-defined macroscopic quantities and are indispensable in the 
Euler relation \eqref{RPSeuler}.}

In the limit $R\rightarrow\infty$, the surface pressure $\hat P\to 0$, therefore 
the pair $(\hat P, \hat A)$ of thermodynamic variables decouples. Moreover, 
we can check that
\begin{equation}
\lim_{R\to\infty} E = M, 
\quad \lim_{R\to\infty} T_R = T_{\mathrm{H}},
\quad \lim_{R\to\infty} \hat{\Phi}_R = \hat\Phi = \frac{\sqrt{G}Q}{Lr_h}, 
\quad \lim_{R\to\infty} \mu _R = \mu.
\end{equation}
Thus the quasi-local RPS formalism automatically falls back to the asymptotic RPS
formalism as $R\to\infty$.

Before ending this subsection, let us mention that Eqs.\eqref{RPS1st} and \eqref{RPSeuler}
implies the following Gibbs-Duhem relation,
\begin{align*}
-S\rd T_R +\hat A\rd \hat P -\hat Q\rd \hat\Phi_R-N\rd \hat \mu_R=0.
\end{align*}
This in turn implies that only three of the four intensive quantities $(T_R, \hat P, 
\hat\Phi_R, \mu_R)$ are independent. Denoting the sequence of 
extensive and intensive variables respectively as
\begin{align}
\mathcal{E}^\alpha &= (S, \hat A, \hat Q, N),\quad 
I_\alpha =(T_R, \hat P, \hat\Phi_R, \mu_R),
\end{align}
we have 
\begin{align}
\det H_{\alpha\beta} = 0, 
\label{detK}
\end{align}
where the matrix $H_{\alpha\beta}$ is defined as
\begin{align}
H_{\alpha\beta} \equiv \pfrac{I_\alpha}{\mathcal{E}^\beta} 
=\ppfrac{E}{\mathcal{E}^\alpha}{ \mathcal{E}^\beta}.
\end{align}
$H_{\alpha\beta}$ is the Hessian of the internal energy $E$ as a thermodynamic potential 
of an open thermodynamic system. All these are standard features of 
extensive thermodynamic systems.

%%%%%%%%%%%%%%%%%%%%%%%%%%%%%%
\subsection{RN-AdS black hole}
\label{sec:RNAdS}
%%%%%%%%%%%%%%%%%%%%%%%%%%%%%%

For RN-AdS black hole,
\begin{equation}
f\left(r\right)=1-\frac{2GM}{r}+\frac{GQ^2}{r^2}+\frac{r^2}{\ell^2},\quad
~f_0\left(r\right)=1+\frac{r^2}{\ell^2}.
\end{equation}
The quasi-local observables described in Eqs.\eqref{QE}-\eqref{RN-Phi-Q} 
should now be changed into 
\blue{
\begin{align}
&E=\frac{R}{G}\left( \sqrt{1+\frac{R^2}{\ell ^2}}-\sqrt{f(R)} \right) ,
\\
&\hat{P}=\frac{1}{8\pi R}\left[ \frac{1}
{\sqrt{f\left( R \right)}}
\left( 1-\frac{GM}{R}+\frac{2R^2}{\ell ^2} \right) 
-\frac{2R^2+\ell ^2}
{\ell ^2\sqrt{1+R^2/\ell ^2}} \right] ,
\quad \hat{A}=\frac{4\pi R^2}{G},
\\
&T_R=\frac{T_{\mathrm{H}}}{\sqrt{f\left( R \right)}}
=\frac{3r_{h}^{4}+r_{h}^{2}\ell ^2-GQ^2\ell ^2}
{4\pi \ell ^2r_{h}^{3}
\sqrt{f\left( R \right)}},
\quad S=\frac{\pi r_{h}^{2}}{G},
\\
&\hat{\Phi}_R=\frac{\sqrt{G}Q}
{L\sqrt{f\left( R \right)}}
\left( \frac{1}{r_h}-\frac{1}{R} \right) ,
\quad\qquad\qquad \hat{Q}=\frac{QL}{\sqrt{G}}.
\end{align}
}
Correspondingly, the quasi-local
chemical potential $\mu_R$ is given as
\begin{equation}
\mu_R=\frac{T_RI_E+\hat{P}\hat{A}}{N}
=\frac{1}{2L^2}\left( \frac{R}{\sqrt{1+R^2/\ell ^2}}
-\frac{R-r_h}{\sqrt{f\left( R \right)}} \right).
\end{equation}
It can be checked that both the first law and the Euler relation 
in the quasi-local RPS formalism remain valid,
\begin{align}
\rd E&=T_R\rd S-\hat{P}\rd \hat{A}
+\hat{\Phi}_R\rd \hat{Q}+\mu_R\rd N,
\\
E&=T_RS-\hat{P}\hat{A}+\hat{\Phi}_R\hat{Q}+\mu_RN.
\end{align}
In the limit $R\to\infty$, the above quasi-local RPS formalism automatically 
falls back to the asymptotic RPS formalism, however only after a Weyl rescaling 
of the boundary metric, as is often done in the context of AdS/CFT correspondence
\cite{Zeyuan:2021uol}.

%%%%%%%%%%%%%%%%%%%%%%%%%%%%%%
\section{RN BHs in the quasi-local RPS formalism}
\label{se:process}
%%%%%%%%%%%%%%%%%%%%%%%%%%%%%%

We have seen in the last section that, for 
RN and RN-AdS BHs, the quasi-local RPS formalism  differ from the 
asymptotic RPS formalism by the inclusion of an extra pair of thermodynamic 
quantities $(\hat P, \hat A)$. Therefore, it is natural to expect that 
the thermodynamic behavior of these BHs will be different in the quasi-local 
and asymptotic RPS formalisms. However, detailed analysis indicates that 
the behavior of RN-AdS BHs in the quasi-local RPS formalism differs 
only quantitatively from the asymptotic case with slightly shifted critical 
parameters, whereas the behavior of RN BHs in the quasi-local RPS formalism differs
qualitatively from the asymptotic case, which is more significant. 
Therefore, in this section, we will \blue{present} the analysis for the pure RN BHs 
\blue{ in great detail and leave a brief description for the RN-AdS BHs case
to the next section}.

Before proceeding, let us mention that, \blue{for the 
pure RN BHs in the asymptotic description, the isocharge $T_{\rm H}-S$ relation 
has only a single maximum, therefore no first order phase equilibrium could happen.}
However, we shall see that the same BHs 
in the quasi-local RPS formalism have more interesting and complicated behaviors.

\subsection{Euler homogeneity}

As mentioned in the introduction, the Euler homogeneity is 
a distinguished feature of the RPS formalism. In the 
quasi-local RPS formalism for RN BHs, the Euler homogeneity is guaranteed 
by the Euler relation \eqref{RPSeuler}. In this subsection, we will 
proceed one step further by rewriting the internal energy $E$ as an explicit 
first order homogeneous function and 
$T_R, ~\hat P, ~\hat \Phi_R, ~\mu_R$ as explicit zeroth order homogeneous functions
in the extensive variables $(S, \hat{A}, \hat{Q}, N)$. 

Using the definition of $S,~\hat{A},~\hat{Q}$
in Eqs.~\eqref{RN-P-A},~\eqref{RN-T-S}
and \eqref{RN-Phi-Q} together with the definition $N=L^2/G$ for $N$,
one can express $R,~r_h,~G,~Q$ as
\begin{equation}
R=\frac{L}{2}\sqrt{\frac{\hat{A}}{\pi N}},
\quad r_h=L\sqrt{\frac{S}{\pi N}},
\quad G=\frac{L^2}{N},
\quad Q=\frac{\hat{Q}}{\sqrt{N}}.
\label{ASNQ}
\end{equation}
Therefore, the above-mentioned thermodynamic functions can be expressed as
\begin{align}
E(S, \hat{A}, \hat{Q}, N)&=\frac{N^{1/2}\hat{A}^{1/2}}{2\pi ^{1/2}L}
\left[ 1-\sqrt{\left( 1-2\sqrt{\frac{S}{\hat{A}}} \right) 
\left( 1- \frac{2\pi \hat{Q}^2}
{\hat{A}^{1/2}S^{1/2}N} \right)} ~\right],
\label{RN-E}
\\
T_R(S, \hat{A}, \hat{Q}, N)&=\frac{\hat{A}^{1/2}\left( SN-\pi \hat{Q}^2 \right)}
{4L\pi ^{1/2}N^{1/2}S^{5/4}\sqrt{
\left( \hat{A}^{1/2}S^{1/2}N-2\pi \hat{Q}^2 \right) 
\left( \hat{A}^{1/2}-2S^{1/2} \right)}},
\label{RN-T}
\\
\hat{P}(S, \hat{A}, \hat{Q}, N)&=\frac{N^{1/2}}{4\pi ^{1/2}L\hat{A}^{1/2}}
\left[ \frac{\hat{A}^{1/2}S^{1/2}N-SN-\pi \hat{Q}^2}
{S^{1/4}N^{1/2}\sqrt{
\left( \hat{A}^{1/2}S^{1/2}N-2\pi \hat{Q}^2 \right) 
\left( \hat{A}^{1/2}-2S^{1/2} \right)}}
-1 \right],
\label{RN-P}
\\
\hat{\Phi}_R(S, \hat{A}, \hat{Q}, N)&=\frac{\pi ^{1/2}\hat{Q}
\left( \hat{A}^{1/2}-2S^{1/2} \right) ^{1/2}}
{LS^{1/4}\sqrt{\hat{A}^{1/2}S^{1/2}N-2\pi \hat{Q}^2}},
\label{RN-Phi}
\\
\mu _R(S, \hat{A}, \hat{Q}, N)&=\frac{\hat{A}^{1/2}}{4\pi ^{1/2}LN^{1/2}}
\left[ 1-\frac{S^{1/4}N^{1/2}
\left( \hat{A}^{1/2}-2S^{1/2} \right) ^{1/2}}
{\sqrt{\hat{A}^{1/2}S^{1/2}N-2\pi \hat{Q}^2}} \right].
\end{align}
If $S,~\hat{A},~\hat{Q},~N$ are rescaled as
$S\rightarrow\lambda S,
~\hat{A}\rightarrow\lambda\hat{A},
~\hat{Q}\rightarrow\lambda\hat{Q},
~N\rightarrow\lambda N$,
$E$ will be rescaled as $E\rightarrow\lambda E$
whereas $T_R,~\hat{P},~\hat{\Phi}_R,~\mu_R$
will not be rescaled.
Therefore, the first-order homogeneity of $E$
and the zeroth order homogeneity of 
$T_R,~\hat{P},~\hat{\Phi}_R,~\mu_R$ are guaranteed.
Please be reminded that, in order to maintain the positivity of the 
Tolman temperature $T_R$ and keep the observers to be located outside the 
event horizon, we have two natural constraint conditions 
over the extensive variables, {\em i.e.} 
$SN>\pi\hat{Q}^2$ and $\hat{A}>4S$.

\subsection{Stability analysis}

We have seen that the quasi-local RPS formalism for RN BHs has all the standard 
features of extensive thermodynamic systems. This allows us to analyze the 
thermodynamic stability of RN BHs using the procedures as described in
any text books like \cite{1987Thermodynamics}. For practical reasons, we
are particularly interested in the stability of RN BHs considered as closed 
thermodynamic systems, {\em i.e.} with fixed $N$ (and therefore fixed 
Newton constant. With this constraint imposed, the sequence of extensive variables 
becomes
\begin{align}
\tilde {\mathcal{E}}^\alpha = (S, \hat A, \hat Q),
\end{align}
and we can use the internal energy criterion to obtain the stability conditions 
for RN BHs. The required stability conditions are
\begin{align}
\pfrac{^2E}{S^2}_{\hat{A},\hat{Q},N}>0,\quad
\pfrac{^2E}{\hat{A}^2}_{S,\hat{Q},N}>0,\quad
\pfrac{^2E}{\hat{Q}^2}_{S,\hat{A},N}>0,
\label{stabilitycond}
\end{align}
which are equivalent to the positivity of the following three thermodynamic 
parameters,
\begin{align}
C_{\hat{A},\hat{Q},N} \equiv T_R \pfrac{S}{T_R}_{\hat{A},\hat{Q},N},\quad
\kappa_{S,\hat{Q},N} \equiv - \frac{1}{\hat A}
\pfrac{\hat A}{\hat P}_{S,\hat{Q},N},\quad 
\eta_{S,\hat{A},N} \equiv \frac{1}{\hat Q}
\pfrac{\hat Q}{\hat \Phi_R}_{S,\hat{A},N}.
\end{align}
An alternative description for these stability conditions can be given by the 
strict positivity of the determinant of the reduced Hessian
\begin{align}
\tilde H_{\alpha\beta} =\ppfrac{E}{\tilde {\mathcal{E}}^\alpha}
{\tilde {\mathcal{E}}^\beta}
\end{align}
as well as all of its main cofactors, {\em i.e.} the algebraic cofactor of the 
diagonal elements of $\tilde H_{\alpha\beta}$. 
In the present case, using Eq.\eqref{RN-E}, we have
\begin{align}
\pfrac{^2E}{S^2}_{\hat{A},\hat{Q},N}&=
\hat{A}^{1/2}\frac{\hat{A}^{1/2}
\left( 3S^2N^2-6SN\pi \hat{Q}^2-5\pi ^2\hat{Q}^4 \right) 
-S^{1/2}\left( \hat{A}N+4\pi \hat{Q}^2 \right)
\left( SN-3\pi \hat{Q}^2 \right)}
{8L\pi ^{1/2}S^{9/4}
\left( S^{1/2}\hat{A}^{1/2}N-2\pi \hat{Q}^2 \right) ^{3/2}
\left( \hat{A}^{1/2}-2S^{1/2} \right) ^{3/2}},
\\
\pfrac{^2E}{\hat{A}^2}_{S,\hat{Q},N}&=
S^{1/4}N\frac{S^{1/2}\hat{A}^{1/2}N
\left( \hat{A}-3S^{1/2}\hat{A}^{1/2}+3S \right) 
-\left( 3\hat{A}-6\hat{A}^{1/2}S^{1/2}+4S \right) 
\pi \hat{Q}^2}
{8L\pi ^{1/2}A^{3/2}
\left( S^{1/2}\hat{A}^{1/2}N-2\pi \hat{Q}^2 \right) ^{3/2}
\left( \hat{A}^{1/2}-2S^{1/2} \right) ^{3/2}}\nonumber
\\
&+\frac{\left( 3\hat{A}^{1/2}-4S^{1/2} \right)
\pi ^{3/2}\hat{Q}^4}
{8LA^{3/2}S^{1/4}
\left( S^{1/2}\hat{A}^{1/2}N-2\pi \hat{Q}^2 \right) ^{3/2}\left( \hat{A}^{1/2}-2S^{1/2} \right) ^{3/2}}
-\frac{N^{1/2}}{8\pi ^{1/2}L\hat{A}^{3/2}},
\\
\pfrac{^2E}{\hat{Q}^2}_{S,\hat{A},N}&=
\frac{\pi ^{1/2}S^{1/4}\hat{A}^{1/2}N
\left( \hat{A}^{1/2}-2S^{1/2} \right) ^{1/2}}
{L\left( S^{1/2}\hat{A}^{1/2}N-2\pi \hat{Q}^2 \right) ^{3/2}},
\\
\ppfrac{E}{S}{\hat{Q}}_{\hat{A},N}&=
-\frac{\pi ^{1/2}\hat{Q}\hat{A}^{1/2}
\left( S^{1/2}\hat{A}^{1/2}-SN-\pi \hat{Q}^2 \right)}
{2LS^{5/4}
\left( S^{1/2}\hat{A}^{1/2}N-2\pi \hat{Q}^2 \right) ^{3/2}
\left( \hat{A}^{1/2}-2S^{1/2} \right) ^{1/2}},
\\
\ppfrac{E}{S}{\hat{A}}_{\hat{Q},N}&=
-\frac{\left( SN-\pi \hat{Q}^2 \right) 
\left( S\hat{A}^{1/2}N+
\left( \hat{A}^{1/2}-4S^{1/2} \right) \pi \hat{Q}^2 \right)}
{8L\pi ^{1/2}S^{5/4}\hat{A}^{1/2}
\left( S^{1/2}\hat{A}^{1/2}N-2\pi \hat{Q}^2 \right) ^{3/2}
\left( \hat{A}^{1/2}-2S^{1/2} \right) ^{3/2}},
\end{align}
\blue{
\begin{align}
\ppfrac{E}{\hat{Q}}{\hat{A}}_{S, N}
&=\frac{\pi ^{1/2}\hat{Q}\left( SN-\pi \hat{Q}^2 \right)}
{2LS^{1/4}\hat{A}^{1/2}
\left( S^{1/2}\hat{A}^{1/2}N-2\pi \hat{Q}^2 \right) ^{3/2}
\left( \hat{A}^{1/2}-2S^{1/2} \right) ^{1/2}}.
\end{align}
}
It is straightforward to see that the condition $\displaystyle 
\pfrac{^2E}{\hat{Q}^2}_{S,\hat{A},N}>0$ always holds, however, due to the complicated
form of the rest partial derivatives given above, it is hard to 
check whether the first two conditions listed in Eq.\eqref{stabilitycond} 
hold or not by means of analytical method. Using numerical techniques, we can check that 
$\det \tilde H_{\alpha\beta}$ is always negative, indicating that 
the RN BHs viewed as closed thermodynamic systems are not globally stable.
Meanwhile, we can also check that $\displaystyle 
\pfrac{^2E}{\hat{A}^2}_{S,\hat{Q},N}$ is always 
positive. Therefore, the only possible instability comes from 
the non-global positivity of $\displaystyle \pfrac{^2E}{S^2}_{\hat{A},\hat{Q},N}$, 
or equivalently from the non-monotonicity of the $T_R-S$ relation. 
In the next subsection, we shall see that such instabilities occur only locally 
in the space of macroscopic states, and the occurrence of such instabilities
indicates that there can be $T_R-S$ phase transitions in RN BHs, even though
such BHs are asymptotically flat. This makes a significant difference 
between the quasi-local and the asymptotic RPS formalisms.

\subsection{$T_R-S$ phase transitions}
\label{sec:T-S-RN}

The non-monotonicity of the $T_R-S$ relation occurs if the partial derivative 
$\displaystyle \pfrac{T_R}{S}_{\hat{Q},\hat{A},N}$ has zero(s). It so happens that 
this partial derivative can have infinitely many zeros. Moreover, 
in the physical range $(SN>\pi\hat{Q},~\hat{A}>4S)$ of parameters, 
$\displaystyle \pfrac{T_R}{S}_{\hat{A},\hat{Q},N}$ and 
$\displaystyle \pfrac{^2T_R}{S^2}_{\hat{A},\hat{Q},N}$ have \blue{a one-parameter
family of} common zeros parametrized by the following algebraic equations,
\begin{equation}
S=\frac{5\pi\hat{Q}^2}{N},
\quad \hat{A}=4\left(2+\sqrt{5}\right)^2\frac{\pi\hat{Q}^2}{N}.
\label{RN-SA}
\end{equation}
For fixed $N$, the above system of equations corresponds to a 
one-dimensional curve in the space $\mathrm{span}(\tilde{\mathcal{E}}^\alpha)$. 
Each point on this curve is a saddle point on the
three-dimensional hypersurface described by the function $T_R(S, \hat A,\hat Q)$ 
as given in Eq.\eqref{RN-T}.
\blue{For static observers at different radial distances,
the critical value of $S,\hat{Q}$ can be different,
which reveals the observer-dependent effect.}
Therefore, the $T_R-S$ phase transitions in RN BHs
in the quasi-local RPS formalism 
can be very complicated. 

To make things simpler, let us recall a crucial physical setting, {\em i.e.} 
the observers are static and located at fixed radial coordinate $r=R$. 
This implies that $\hat A$ needs to be fixed
\blue{to neglect the influence of different choice of $\hat{A}$}.
In this case, 
$\displaystyle \pfrac{T_R}{S}_{\hat{A},\hat{Q},N}$ and 
$\displaystyle \pfrac{^2T_R}{S^2}_{\hat{A},\hat{Q},N}$ have only a single 
common zero, which is a critical point for the $T_R-S$ phase transition. 
The critical values of $S, \hat{Q}$ and $T_R$ read
\begin{align}
S_c=\frac{5\hat{A}}{4\left( 2+\sqrt{5} \right) ^2},\quad
\hat{Q}_c=\frac{\hat{A}^{1/2}N^{1/2}}{2\left( 2+\sqrt{5} \right) \pi ^{1/2}},\quad
T_c=\frac{\left(2+ \sqrt{5} \right) ^{3/2} N^{1/2}}
{5^{5/4}\pi^{1/2} L \hat{A}^{1/2}}.
\end{align}

In order to describe the $T_R-S$ phase transition clearly, we also need to 
introduce the Helmholtz free energy
\begin{equation}
F_R\left( T_R,\hat{A},\hat{Q},N \right) =E\left( S,\hat{A},\hat{Q},N \right) -T_R S
\end{equation}
and the isocharge heat capacity
\begin{align}
C_{\hat{A},\hat{Q},N}&=T_R\pfrac{S}{T_R}_{\hat{A},\hat{Q},N}
\nonumber\\
&=\frac{2S\left( SN-\pi \hat{Q}^2 \right)
\left[ S^{1/2}\left( \hat{A}N+4\pi \hat{Q}^2 \right) -2\hat{A}^{1/2}\left( SN+\pi \hat{Q}^2 \right) \right]}
{\hat{A}^{1/2}\left( 3S^2N^2-6\pi \hat{Q}^2SN-5\pi ^2\hat{Q}^4 \right) 
-S^{1/2}\left( \hat{A}N+4\pi \hat{Q}^2 \right) \left( SN-3\pi \hat{Q}^2 \right)}.
\label{RN-C-QAN}
\end{align}

Using the critical parameters, we can introduce the dimensionless parameters
\begin{align}
s&=\frac{S}{S_c},\quad q=\frac{\hat{Q}}{\hat{Q}_c}.
\end{align}
The $T_R-S$ relation, the Helmholtz free energy and the isocharge 
heat capacity can all be reparametrized using the dimensionless parameters, yielding
\begin{align}
t&\equiv \frac{T_R}{T_c}=\frac{5s-q^2}{4s^{3/2}}
\sqrt{\frac{4\left( \sqrt{5}+2 \right) s^{1/2}}
{\sqrt{5}s^{1/2}\left( q^2+\left( 2+\sqrt{5} \right) ^2 \right) 
-\left( 2+\sqrt{5} \right) \left( 5s+q^2 \right)}},
\label{RN-t-sq}
\\
\mathcal{F}&\equiv \frac{F_R}{F_c}=\frac{2\left( 2+\sqrt{5} \right) ^{1/2}}
{2\left( 2+\sqrt{5} \right) ^{1/2}-5^{3/4}}
\left[ 1-\frac{4s^{1/2}\left( q^2+\left( 2+\sqrt{5} \right) ^2 \right) 
-\sqrt{5}\left( 2+\sqrt{5} \right) \left( 3s+q^2 \right)}
{4\left( 2+\sqrt{5} \right) ^2s^{1/2}\sqrt{f\left( s,q \right)} } \right] ,
\label{RN-f-sq}
\end{align}
with
\begin{equation}
f\left( s,q \right) =\frac{\sqrt{5}s^{1/2}\left( q^2+\left( 2+\sqrt{5} \right) ^2 \right)
 -\left( 2+\sqrt{5} \right) \left( 5s+q^2 \right)}
{\sqrt{5}\left( 2+\sqrt{5} \right) ^2s^{1/2}},
\end{equation}
and
\begin{equation}
c_{\hat{Q},\hat{A},N} \equiv 
\frac{C_{\hat{Q},\hat{A},N}}{\blue{c_1}}=\frac{s\left( 5s-q^2 \right)
 \left[ \sqrt{5}s^{1/2}\left( \left( \sqrt{5}+2 \right) ^2+q^2 \right) 
-\left( 2+\sqrt{5} \right) \left( 5s+q^2 \right) \right]}
{\sqrt{5}\left( 2+\sqrt{5} \right) \left( 15s^2-6sq^2-q^4 \right) 
-2s^{1/2}\left( \left( 2+\sqrt{5} \right) ^2+q^2 \right) \left( 5s-3q^2 \right)},
\end{equation}
with $\blue{c_1}=\sqrt{5}\hat{A}/\left(2+\sqrt{5}\right)^2$.
In the limit $q\rightarrow 0$, the RN BHs degenerate into Schwarzschild BHs,
and we have the following limit behavior,
\blue{
\begin{align}
t&\rightarrow \frac{5^{3/4}}{2s^{1/2}\sqrt{2+\sqrt{5}-\sqrt{5}s^{1/2}}},
\end{align}
\begin{align}
\mathcal{F} &\rightarrow \frac{2\left( 2+\sqrt{5} \right) ^{1/2}}
{2\left( 2+\sqrt{5} \right) ^{1/2}-5^{3/4}}
\left( 1-\frac{4\left( 2+\sqrt{5} \right) -3\sqrt{5}s^{1/2}}
{4\left( 2+\sqrt{5} \right) ^{1/2}\sqrt{2+\sqrt{5}-\sqrt{5}s^{1/2}}} \right) ,\\
c_{\hat{Q},\hat{A},N} &\rightarrow 
\frac{\sqrt{5}s\left( 2+\sqrt{5}-\sqrt{5}s^{1/2} \right)}
{3\sqrt{5}s^{1/2}-2\left( 2+\sqrt{5} \right)}.
\label{RN-t-q0}
\end{align}
}

\begin{figure}[htbp]
  \centering
  \begin{subfigure}[b]{0.45\textwidth}
    \centering
    \includegraphics[width=\textwidth]{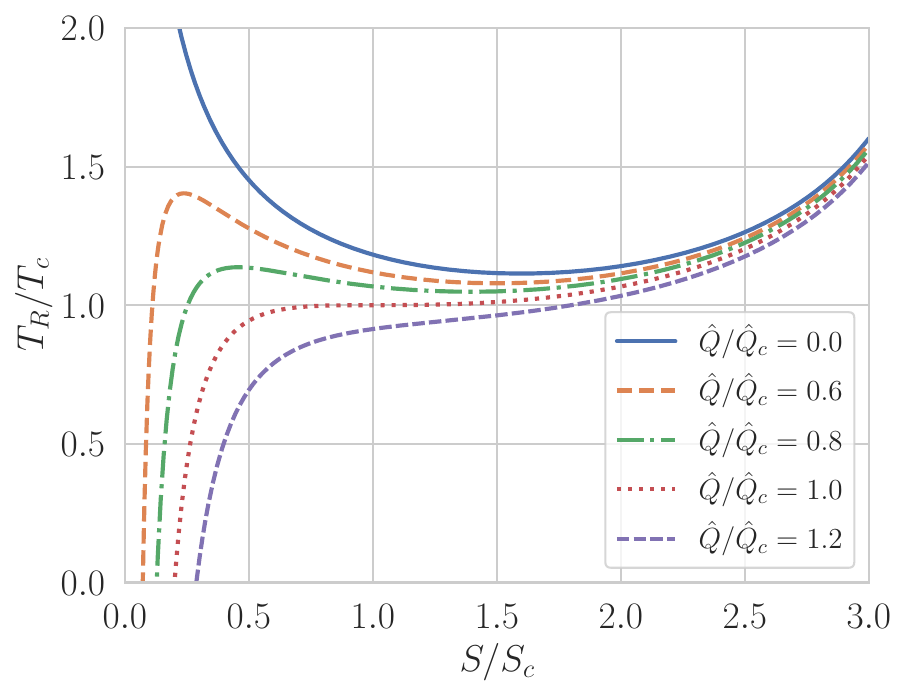}
  \end{subfigure}
  \hfill
  \begin{subfigure}[b]{0.45\textwidth}
    \centering
    \includegraphics[width=\textwidth]{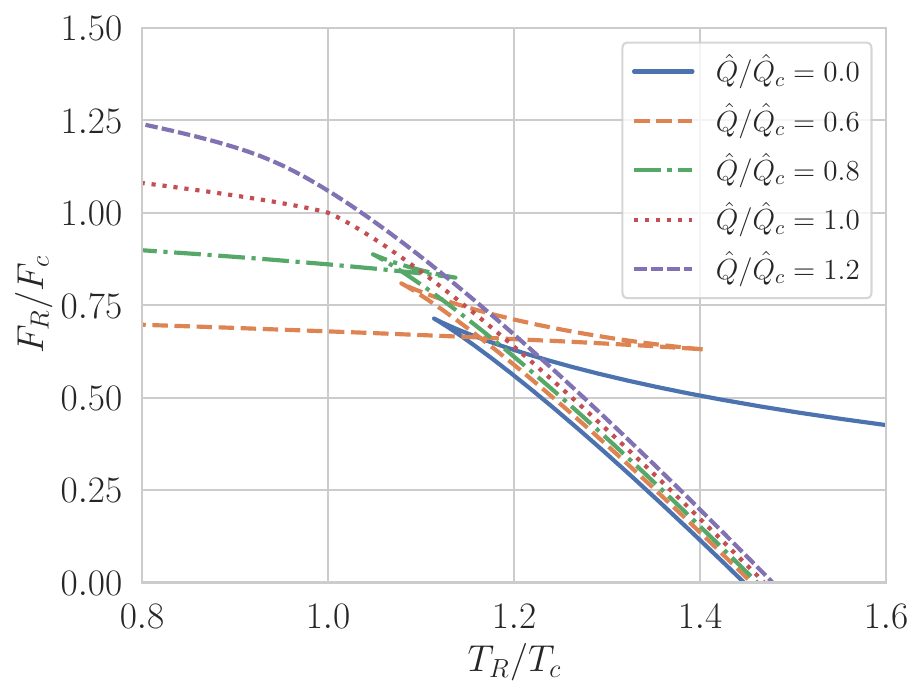}
  \end{subfigure}
  \caption{$T_R-S$ and $F_R-T_R$ curves in the isocharge processes}
  \label{fig:RN-TS-fixQ}
\end{figure}

\begin{figure}[htbp]
    \centering
    \includegraphics[width=0.45\textwidth]{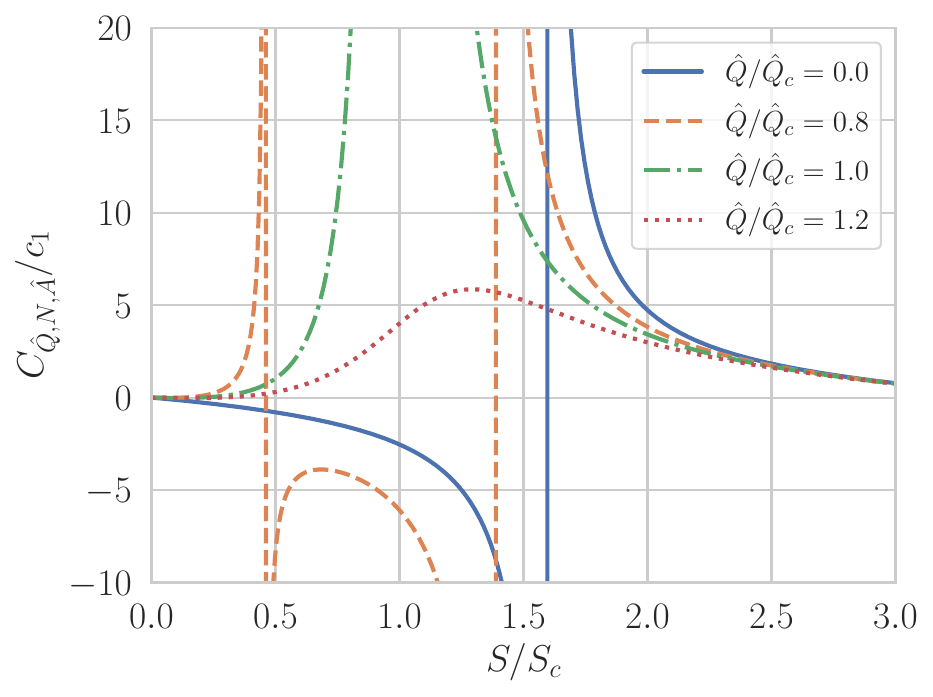}
    \caption{heat capacity
    curves in the isocharge processes}
    \label{fig:RN-Cq-Q}
\end{figure}

Although the above analytical results are exact and sufficient to analyze
the $T_R-S$ transitions, it is more illustrative to depict the corresponding 
isocharge $T_R-S$ and $F_R-T_R$ curves. This is done using the python package matplotlib 
and the resulting curves are presented in Fig.\ref{fig:RN-TS-fixQ}.
Correspondingly, the isocharge $C_{\hat{Q},\hat{A},N}-S$ curves are depicted as 
in Fig.\ref{fig:RN-Cq-Q}.

From the $T_R-S$ curves one can see that, there are two stable phases at super
critical temperatures and non-zero subcritical charges. The swallow tail shape of the 
$F_R-T_R$ curves indicates that, at super critical temperatures, 
the $T_R-S$ phase transitions are of the first order, while at the 
critical temperature, it becomes the second order. The 
$C_{\hat{Q},\hat{A},N}-S$ curves further confirms the above judgements.
Let us stress that all these results are based on placing the observers at a
fixed and finite radial distance. \blue{The high entropy and high temperature 
stable phase arises from the fact that when the event horizon approaches the 
place where the observers reside, the redshift factor in Eq.\eqref{TR215} increases
significantly. This phenomenon could not happen 
if the observers are placed at the radial infinity. Moreover, 
for observers at radial infinity, }
$\hat A$ becomes divergent and hence $S_c$ and $\hat Q_c$ also diverge, whereas
$T_c$ goes to zero. Therefore, it is not surprising that no \blue{first order phase 
equilibrium} could be observed in the asymptotic RPS formalism for the RN BHs.

\subsection{Isovoltage processes}

Although the isocharge $T_R-S$ phase transition always occurs when the 
observers are placed at finite radial distance, it is not the case in the 
isovoltage processes. This is illustrated in the follows. 
Using Eqs.~\eqref{RN-T} and \eqref{RN-Phi}, 
the Tolman temperature can be reparametrized in terms of $(S, \hat A, \hat\Phi_R)$ 
instead of $(S, \hat A, \hat Q)$. The result reads 
\begin{equation}
T_R=\frac{N^{1/2}\hat{A}^{1/4}
\left( 1-L^2\hat{\Phi}_{R}^{2} \right)}
{4\pi ^{1/2}LS^{1/2}
\sqrt{\hat{A}^{1/2}-2S^{1/2}
\left( 1-L^2\hat{\Phi}_{R}^{2} \right)}}.
\end{equation}

\begin{figure}[h]
    \centering
    \includegraphics[width=0.45\textwidth]{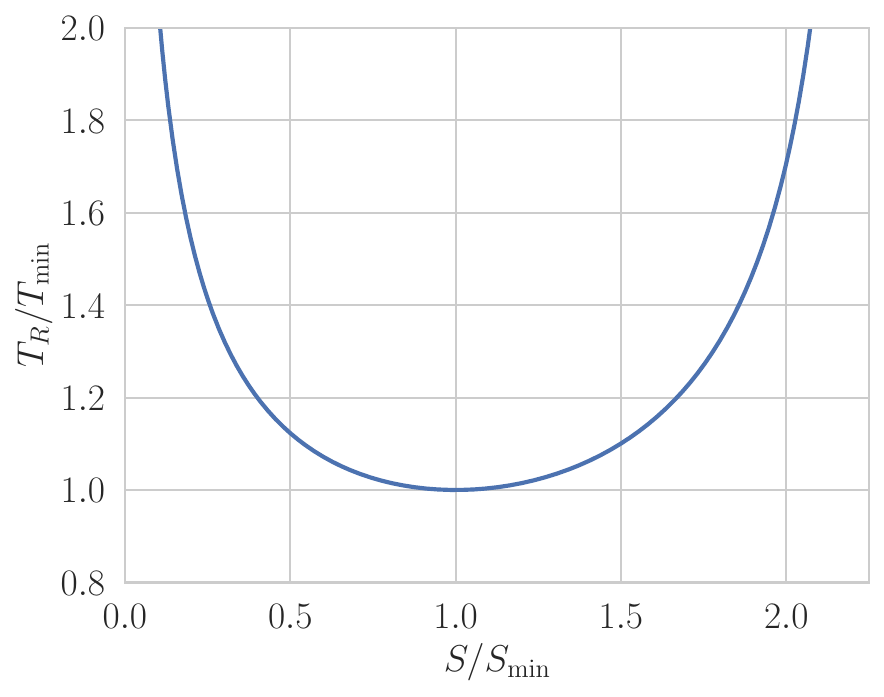}
    \caption{$T_R-S$
    curves in the iso-voltage processes}
    \label{fig:T-S-Phi}
\end{figure}

At fixed $N, \hat A$ and $\hat\Phi_R$, the $T_R-S$ curve possesses a single 
minimum located at
\begin{equation}
S_{\min}=\frac{\hat{A}}
{9\left(1-L^2\hat{\Phi}_{R}^{2}\right)^2},
\quad T_{\min}=\frac{3\sqrt{3}N^{1/2}
\left( 1-L^2\hat{\Phi}_{R}^{2} \right) ^2}
{4\pi ^{1/2}L\hat{A}^{1/2}}.
\end{equation}
Thus we can introduce the novel dimensionless parameters 
\begin{equation}
\tilde{s}=\frac{S}{S_{\min}},
\quad \tilde{t}=\frac{T_R}{T_{\min}}
\end{equation}
and rewrite the isovoltage $T_R-S$ relation as 
\begin{equation}
\tilde{t}=\frac{1}{\tilde{s}^{1/2}\sqrt{3-2\tilde{s}^{1/2}}}.
\end{equation}
The corresponding curve is depicted in Fig.\ref{fig:T-S-Phi}. It turns out that 
there is only a single stable phase with $S>S_{\min}$. The $S<S_{\min}$ branch is 
thermodynamically unstable. Therefore there can be no phase equilibrium 
in the isovoltage processes. Similar isovoltage $T-S$ behavior was found 
in the asymptotic RPS formalism for the RN-AdS BHs \cite{Zeyuan:2021uol}.

\subsection{Schwarzschild limit and Hawking-Page-like transitions}

In this subsection, we consider the neutral limit by setting $\hat Q=0$. 
This is nothing but the case of Schwarzschild BHs. In this limiting case, 
the thermodynamic potential $\Psi_R$ defined in Eq.\eqref{PsiR} degenerates 
with the Helmholtz free energy, thus we have 
\begin{align*}
F_R=T_R I_E,
\end{align*}
with 
\begin{equation}
I_E=\frac{4\pi r_hR}{G} \left( \sqrt{1-\frac{r_h}{R}}-1 +\frac{3r_h}{4R} \right) .
\end{equation}
If $R\to\infty$, $I_E$ regarded as a function in $r_h$ 
has no zeros. However, if $R$ remains finite and 
$r_h =0$ or $r_h =8R/9$, $I_E$ becomes zero. 

Let us recall that, in the asymptotic description of Schwarzschild-AdS BHs, 
the zero of $I_E$ plays a very special role: it corresponds to the transition 
from neutral BHs to relativistic thermal gases known as the Hawking-Page 
phase transition \cite{Hawking:1982dh}. For asymptotically flat Schwarzschild BHs, 
the absence of zeros of $I_E$ when $R\to \infty$ indicates that there can be no 
Hawking-Page transitions for the Schwarzschild BHs in the asymptotic description.
However, as is shown above, $I_E$ does have zeros if $R$ is taken to be finite. 
This seems to indicate that some Hawking-Page-like transitions may indeed happen 
for the Schwarzschild BHs in the quasi-local description. 

The two zeros of $I_E$ separates the physical region of $r_h$ into two subregions: 
$\displaystyle 0<r_h<\frac{8R}{9}$ and $\displaystyle \frac{8R}{9}<r_h<R$. 
In the first subregion, $I_E$ is positive, 
and in the second subregion, $I_E$ is negative. The existence of the two zeros 
implies that the function $I_E(r_h)$ has extrema/extremum in the region 
$\displaystyle 0<r_h<\frac{8R}{9}$. It is not difficult to see that the 
only extremum is a maximum located at $\displaystyle r_h = \frac{2R}{3}$, and that 
$\displaystyle \frac{\rd I_E}{\rd r_h}>0$ for $\displaystyle 0< r_h < \frac{2R}{3}$, 
$\displaystyle \frac{\rd I_E}{\rd r_h}<0$ for $\displaystyle \frac{2R}{3}< r_h < R$.
Therefore, if $r_h$ were initially set in the region $\displaystyle  
r_h\in \left(0, \frac{2R}{3}\right)$ and received a small perturbation 
$\delta r_h$, then the Helmholtz free energy increases if $\delta r_h>0$ and decreases
if $\delta r_h<0$. Since the Helmholtz free energy always tends to acquire a 
smaller value in spontaneous thermodynamic processes, we conclude that the 
Schwarzschild BHs with $\displaystyle r_h\in \left(0, \frac{2R}{3}\right)$ 
will always tend to evolve into the $r_h=0$ state, {\em i.e.} a thermal gas state
after the perturbation. On the contrary, if $r_h$ were initially set in the 
region $\displaystyle r_h\in \left(\frac{2R}{3}, R\right)$ and received 
a small perturbation $\delta r_h$, then the Schwarzschild BHs will always 
tend to evolve into a state with bigger $r_h$ after the perturbation. 
All these behaviors are extremely similar to that of the Schwarzschild-AdS BHs in the 
asymptotic description.
\blue{This can serve as an evidence for Hawking-Page-like transition.}

%%%%%%%%%%%%%%%%%%%%%%%%%%%%%%
\section{\blue{RN-AdS BHs in the quasi-local RPS formalism}}
\label{se:process-AdS}
%%%%%%%%%%%%%%%%%%%%%%%%%%%%%%

\blue{In this section, we will give a brief discussion about the thermodynamic 
behavior of RN-AdS BHs.
Let us recall that the thermodynamics for the RN-AdS BHs in the asymptotic RPS formalism 
has been analyzed in detail in Ref.\cite{Zeyuan:2021uol}, which predicts that
a first order phase transition could appear at supercritical temperature 
and subcritical value of $\hat{Q}$ in the isocharge processes, and that 
the isovoltage $T_{\mathrm{H}}-S$ curve has a single minimum,
which signifies a non-equilibrium phase transition. We will see that these behaviors 
are inherited also in the quasi-local description, with only quantitative differences 
in the critical parameters. }

\blue{For AdS BHs, we are free to identify the characteristic length scale $L$ with 
the AdS radius $\ell$. If we choose $L=\ell$, with the aid of Eq.~\eqref{ASNQ}, 
the thermodynamic functions for the RN-AdS BHs are worked out to be}
\blue{
\begin{align}
E&=\frac{\sqrt{N\hat{A}}}{2\pi ^{1/2}\ell}\left( \sqrt{1+\frac{\hat{A}}{4\pi N}}
-\sqrt{\frac{S^{1 / 2} \left(4  \hat{A} \pi N + 16 \pi^2 \hat{Q}^2  + \hat{A}^2 \right) 
- 8 \hat{A}^{1 / 2} \left(  S \pi N + \pi^2 \hat{Q}^2  +  S^2 \right)}
{4 \hat{A} S^{1 / 2} \pi N}} \right) ,
\\
T_R&=\frac{\hat{A}^{1/2}\left( 3S^2+\pi NS-\pi ^2\hat{Q}^2 \right)}
{2\pi \ell S^{5/4}\sqrt{\left( \hat{A}^{1/2}-2S^{1/2} \right) 
\left( S^{1/2}\hat{A}^{1/2}\left( \hat{A}+2\hat{A}^{1/2}S^{1/2}+4S+4\pi N \right) 
-8\pi ^2\hat{Q}^2 \right)}},
\\
\hat{P}&=\frac{2S^{1/2}\pi N\left( \hat{A}^{1/2}-S^{1/2} \right) -2\pi ^2\hat{Q}^2
+S^{1/2}\left( \hat{A}^{3/2}-2S^{3/2} \right)}
{4\hat{A}^{1/2}S^{1/4}\pi \ell 
\sqrt{\left( \hat{A}^{1/2}-2S^{1/2} \right) 
\left( S^{1/2}\hat{A}^{1/2}\left( \hat{A}+2\hat{A}^{1/2}S^{1/2}+4S+4\pi N \right) 
-8\pi ^2\hat{Q}^2 \right)}}
\nonumber
\\
&-\frac{2\pi N+\hat{A}}{4\pi \hat{A}^{1/2}\ell \sqrt{4\pi N+\hat{A}}},
\\
\hat{\Phi}_R&=\frac{2\pi \hat{Q}
\left( \hat{A}^{1/2}-2S^{1/2} \right) ^{1/2}}
{\ell S^{1/4}
\sqrt{S^{1/2}\hat{A}^{1/2}\left( \hat{A}+2\hat{A}^{1/2}S^{1/2}+4S+4\pi N \right)
 -8\pi ^2\hat{Q}^2}},
\\
\mu _R&=\frac{1}{2\ell}\left( \frac{\hat{A}^{1/2}}{\sqrt{4\pi N+\hat{A}}}
-\frac{\hat{A}^{1/2}S^{1/4}
\left( \hat{A}^{1/2}-2S^{1/2} \right) ^{1/2}}
{\sqrt{S^{1/2}\hat{A}^{1/2}
\left( \hat{A}+2\hat{A}^{1/2}S^{1/2}+4S+4\pi N \right) 
-8\pi ^2\hat{Q}^2}} \right),
\end{align}}
\blue{The first-order homogeneity of $E$
and the zeroth order homogeneity of $T_R,~\hat{P},~\hat{\Phi}_R,~\mu_R$ in the 
independent variables $S,~\hat{A},~\hat{Q},~N$ are obvious. }

\blue{Keeping $\hat{A},N$ fixed, the equations $\left( \dfrac{\partial T_R}
{\partial S} \right) _{\hat{Q},\hat{A},N}
=\left( \dfrac{\partial ^2T_R}{\partial S^2} \right) _{\hat{Q},\hat{A},N}=0$ 
have an inflection point solution in the physical range ($T_R>0, \hat{A}>4S$). 
Explicitly, the equations to be solved are presented as follows,}
\blue{
\begin{align}
&S_{c}^{2}-\pi NS_c+5\pi ^2\hat{Q}_{c}^{2}=0,
\label{SQ1}
\\
\frac{S_{c}^{1/2}
\left( \hat{A}^2+4\pi N\hat{A}+16\pi ^2\hat{Q}_{c}^{2} \right)}
{4\hat{A}^{1/2}}
&=\frac{\left( 3S_{c}^{2}+\pi NS_c-\pi ^2\hat{Q}_{c}^{2} \right) ^2}
{-3S_{c}^{2}+\pi NS_c-3\pi ^2\hat{Q}_{c}^{2}}
+2\left( S_{c}^{2}+\pi NS_c+\pi ^2\hat{Q}_{c}^{2} \right),
\label{SQ2}
\end{align}
}
\blue{
in which we have replaced $S$ by $S_c$ and $\hat Q$ by $\hat Q_c$ in order to stress that 
the solution to the above equations corresponds to the critical parameters. 
Thanks to the Euler homogeneity, $S_c, \hat Q_c$ and $\hat A$ share the same scaling 
properties with $N$. Therefore, it is convenient to introduce the new 
scale-independent parameters }
\blue{
\begin{align}
a\equiv \frac{\hat{A}}{N},\quad
s_c \equiv \frac{S_c}{N},\quad
q_c\equiv \frac{\hat Q_c}{N}.
\end{align}}
\blue{
Using these new parameters, Eqs.\eqref{SQ1} and \eqref{SQ2} can be rewritten as}
\blue{
\begin{align}
&s_{c}^{2}-\pi s_c+5\pi ^2q_{c}^{2}=0,
\\
\frac{s_{c}^{1/2}
\left(a^2+4\pi a+16\pi ^2q_{c}^{2} \right)}
{4a^{1/2}}
&=\frac{\left( 3s_{c}^{2}+\pi s_c-\pi ^2q_{c}^{2} \right) ^2}
{-3s_{c}^{2}+\pi s_c-3\pi ^2q_{c}^{2}}
+2\left( s_{c}^{2}+\pi s_c+\pi ^2q_{c}^{2} \right).
\end{align}
}
\blue{
Unlike the situation of asymptotic description, the solution 
to the above inflection point equations cannot be worked out analytically. 
Therefore, to proceed with the 
analysis of phase transitions, we need to employ numerical techniques. }

\blue{
In order to carry out the numerical process, we need to prescribe some fixed values for $a$.
At several discrete values of $a$, the numerical values of $q_c$ and $s_c$ 
are listed in Table \ref{Tab2}.
}
\begin{comment}
Setting $a=1,\pi$ respectively, the numeric values of the critical 
parameters are found to be 
\begin{align}
s_c\approx 0.061694,
\quad 
q_c \approx 0.062052, \quad (a=1),
\\
s_c \approx 0.156745,
\quad
q_c \approx 0.097369, \quad(a=\pi).
\end{align}
For $a\in(1,10)$, numerical results indicate that the best second-order rational polynomial
fit of $s_c(a)$ relation is
\begin{align}
s_c(a)= \frac{0.001427 + 0.070315\,a + 0.001268\,a^2}{0.024186 + 0.892451\,a + 0.042763\,a^2},\quad 1\le a \le 10
\label{eq:sc_2rational_1to10}
\end{align}
with the largest relative error $\le 0.0022\%$,
\begin{align}
s_c \approx \frac{K}{1+\exp\left(\frac{\alpha-a}{\beta}\right)}
\end{align}
with largest error $\le 7.6\times 10^{-6}$, where
\begin{align}
K=\frac{\pi}{6},\quad
\alpha= 16.310,\quad
\beta = 9.742,	
\end{align}
and $q_c$ is related to $s_c$ precisely via
\begin{align}
q_c(s_c)=\frac{1}{\pi}\sqrt{\frac{s_c(\pi-s_c)}{5}}.	
\end{align}
\end{comment}

\blue{
\begin{table}[htbp]
\begin{center}
\begin{tabular}{|c|c|c|c|c|c|c|}
\hline 
$a$ & 1 & 2  & 3 & 4 & 5 & 6 \tabularnewline
\hline 
\hline 
$q_{c}$ & 0.062052 & 0.083757 & 0.096393 & 0.105031 & 0.111413 & 0.116343 \tabularnewline
\hline 
$s_{c}$ & 0.061694 & 0.114357 & 0.153448 & 0.184065 & 0.208867 & 0.229365 \tabularnewline
\hline 
\end{tabular}
\end{center}
\caption{Critical parameters at different choices of $a$}\label{Tab2}
\end{table}
}

\begin{figure}[!htbp]
  \centering
  \begin{subfigure}[b]{0.45\textwidth}
    \centering
    \includegraphics[width=\textwidth]{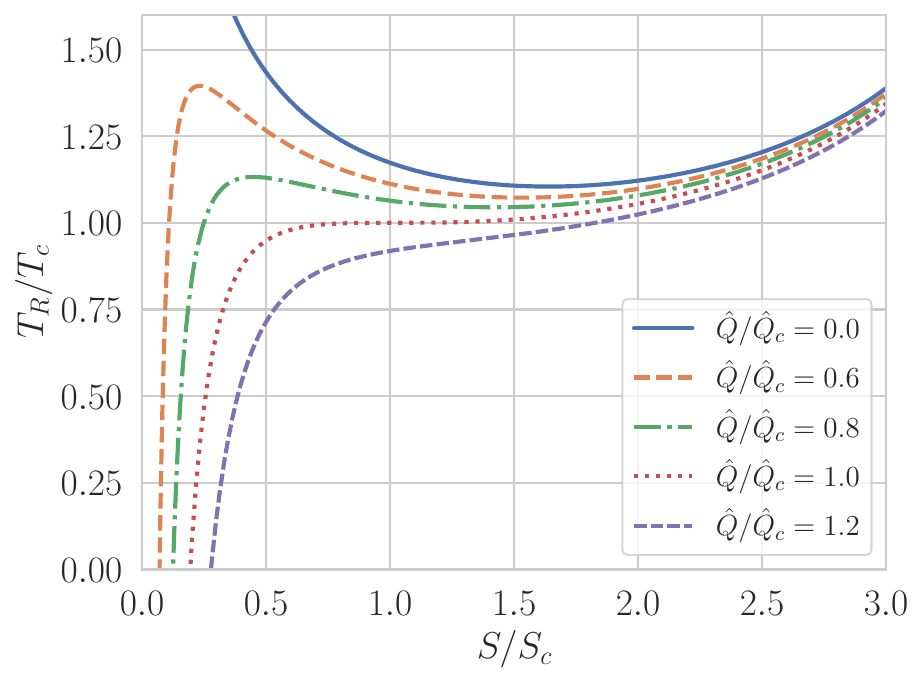}
  \end{subfigure}
  \hfill
  \begin{subfigure}[b]{0.45\textwidth}
    \centering
    \includegraphics[width=\textwidth]{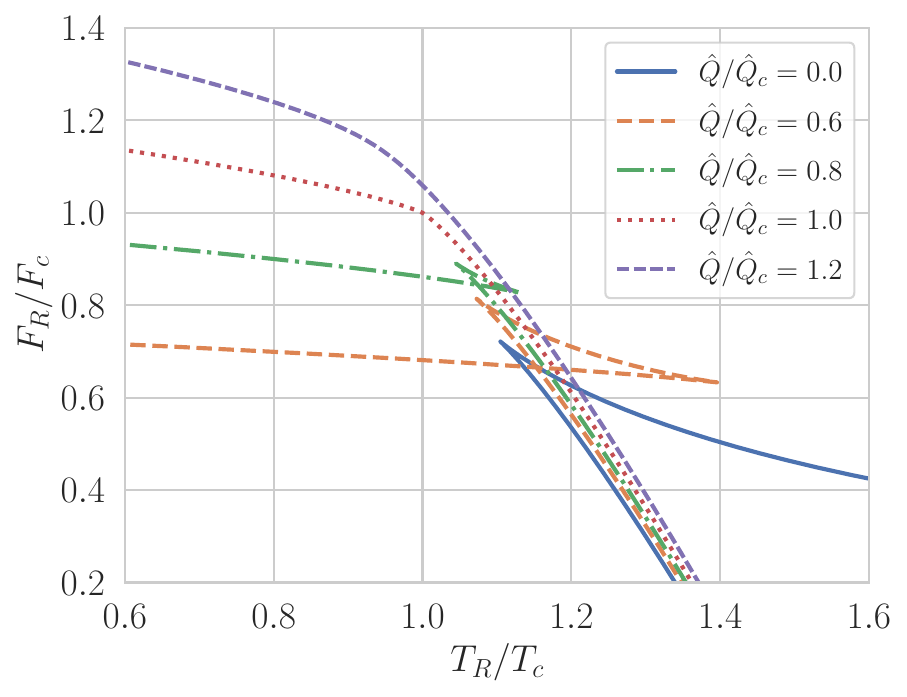}
  \end{subfigure}
  \caption{$T_R-S$ and $F_R-T_R$ curves of RN-AdS BHs in the isocharge processes}
  \label{fig:RNAdS-TS-fixQ}
\end{figure}

\begin{figure}[!htbp]
    \centering
    \includegraphics[width=0.45\textwidth]{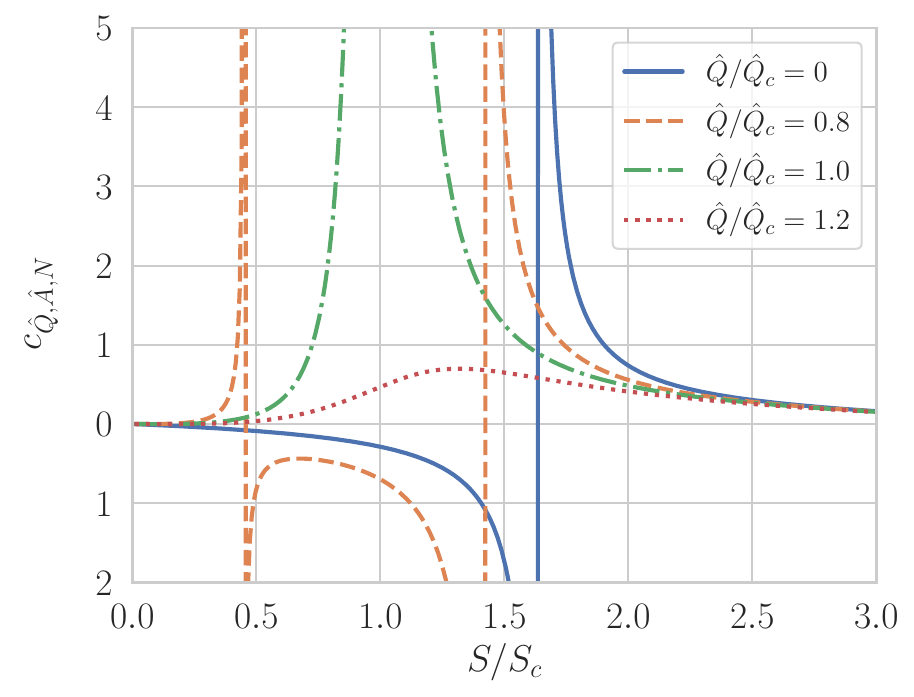}
    \caption{heat capacity
    curves of RN-AdS BHs in the isocharge processes}
    \label{fig:RNAdS-Cq-Q}
\end{figure}

\blue{
As usual, it is convenient to introduce the dimensionless parameters}
\blue{
\begin{align}
{s}& \equiv\frac{S}{S_c},\quad {q} =\frac{\hat{Q}}{\hat{Q}_c},
\end{align}
so that the other thermodynamic quantities can also be described using such parameters,
\begin{align}
{t}& \equiv\frac{T_R}{T_c}=\frac{3s^{2}s^{2}_{c}
+\pi ss_{c}-\pi^{2}q^{2}q^{2}_{c}}{\pi^{1/2}s^{3/2}a^{1/4}s^{1/4}_{c}(-3s^{2}_{c}
+\pi s_{c}-3\pi^{2}q^{2}_{c})^{1/2}\sqrt{f}},
\\
\mathcal{F}& \equiv\frac{F_R}{F_c}
=\frac{1}{f_{0}}\left(\sqrt{1+\frac{a}{4\pi}}-\sqrt{f}-\frac{3s^{2}s_{c}^2
+\pi ss_{c}-\pi^{2}q^{2}q^{2}_{c}}{2\pi a^{1/2}s^{1/2}s^{1/2}_{c}\sqrt{f}}\right),
\\
c_{\hat{Q},\hat{A},N}&\equiv\frac{C_{\hat{Q},\hat{A},N}}{N}
=\frac{8\pi as^{3/2}s^{3/2}_{c}(3s^{2}s^{2}_{c}+\pi ss_{c}
-\pi^{2}q^{2}q^{2}_{c})f}{(\mathcal{C}_{1}+\mathcal{C}_{2})},
\end{align}
}
\blue{
where}
\blue{
\begin{align}
{f} _0&=\sqrt{1+\frac{a}{4\pi}}-\frac{3s^{2}_{c}+3\pi s_{c}
-5\pi^{2}q^{2}_{c}}{2a^{1/4}\pi^{1/2}s^{1/4}_{c}(-3s^{2}_{c}
+\pi s_{c}-3\pi^{2}q^{2}_{c})^{1/2}},
\\
f&=\frac{s^{1/2}s^{1/2}_{c}(4\pi a+a^{2}+16\pi^{2}q^{2}q^{2}_{c})-8a^{1/2}(s^{2}s^{2}_{c}
+\pi ss_{c}+\pi^{2}q^{2}q^{2}_{c})}{4as^{1/2}s^{1/2}_{c}\pi},
\\
\mathcal{C}_{1} & =s^{1/2}s^{1/2}_{c}(3s^{2}s^{2}_{c}
-\pi ss_{c}+3\pi^{2}q^{2}q^{2}_{c})(4\pi a+a^{2}+16\pi^{2}q^{2}q^{2}_{c}),\\
\mathcal{C}_{2} & =4a^{1/2}(3s^{4}s^{4}_{c}+2s^{3}s^{3}_{c}\pi
+3\pi^{2}(1-6q^{2}q_c^2)s^{2}s^{2}_{c}-6\pi^{3}q^{2}q^{2}_{c}ss_{c}-5\pi^{4}q^{4}q^{4}_{c}).
\end{align}
}
\blue{
Using these equations, the $T_R-S$ and $F_R-T_R$ curves of RN-AdS BHs in the 
isocharge processes are depicted in Fig.~\ref{fig:RNAdS-TS-fixQ},
and the isocharge  $C_{\hat{Q},\hat{A},N}-S$ curves 
are depicted in Fig.~\ref{fig:RNAdS-Cq-Q}. While creating these pictures, we have 
taken the choice $a=1$. The curves for other choices of $a$ 
have no qualitative differences from the $a=1$ case, thus we do not bother to 
present the plots of more curves. One can see that 
a first-order phase phase equilibrium occurs at supercritical temperatures
and non-zero subcritical charges. The same behavior also appears in the asymptotic 
description \cite{Zeyuan:2021uol}, and thus for RN-AdS BHs, there is no 
qualitative differences between the quasi-local and asymptotic 
RPS formalisms.}

\begin{figure}[h]
    \centering
    \includegraphics[width=0.45\textwidth]{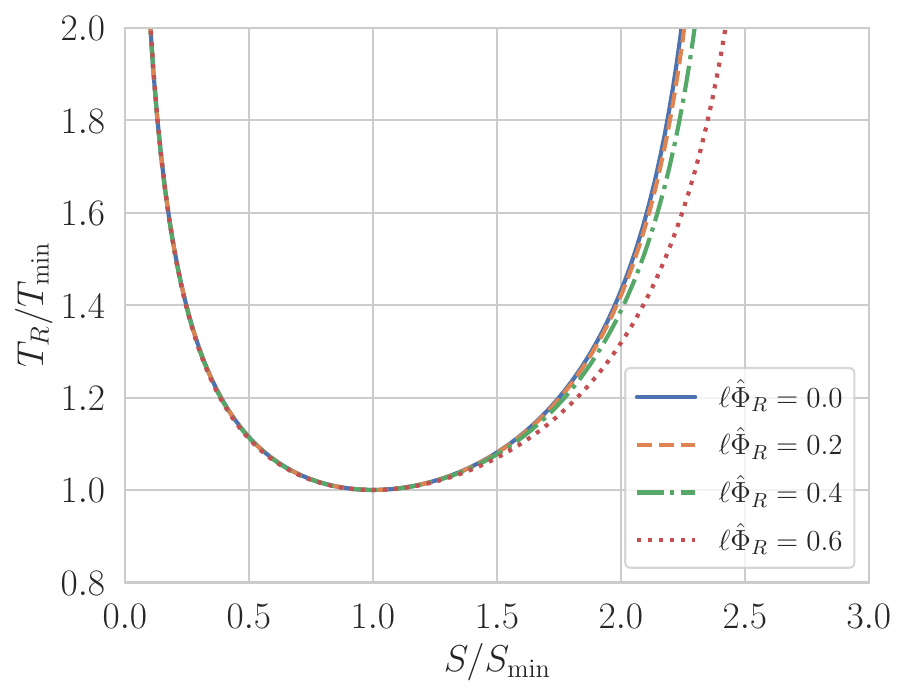}
    \caption{$T_R-S$
    curves of RN-AdS BHs in the iso-voltage processes}
    \label{fig:RNAdS-T-S-Phi}
\end{figure}

\blue{
It remains to consider the isovoltage $T_R-S$ curve for RN-AdS BHs in the quasi-local 
RPS formalism. In order to describe the isovoltage $T_R-S$ curve, we need to rewrite $T_R$
as a function in $(S,\hat\Phi_R,\hat A, N)$. The result reads}
\blue{
\begin{align}
T_R& =\hat{A}^{1/4}\frac{4(3S+\pi N)(1-\ell^{2}\hat{\Phi}^{2}_{R})
-\hat{A}^{1/2}\ell^{2}\hat{\Phi}^{2}_{R}(A^{1/2}+4S^{1/2})}
{8\pi\ell\sqrt{S(\hat{A}^{1/2}-2S^{1/2}(1-\ell^{2}\hat{\Phi}^{2}_{R}))
(\hat{A}+2\hat{A}^{1/2}S^{1/2}+4S+4\pi N)}}	.
\end{align}}
\blue{
Keeping $\hat A, N$ fixed, we find that for any physically admissible value of 
$\hat\Phi_R$, the isovoltage $T_R-S$ curve has only a single minimum. Omitting 
the detailed process, we present only the resulting isovoltage $T_R-S$ curves 
for several choices of $\hat\Phi_R$ in Fig.\ref{fig:RNAdS-T-S-Phi}. 
It appears that, at different $\hat\Phi_R$ values, the isovoltage $T_R-S$ curves 
are also different, which makes a quantitative contrast to the case of RN BHs 
(see Fig.\ref{fig:T-S-Phi}) as well as all previously known cases of AdS BHs 
in the asymptotic RPS formalism. However, on the qualitative level, 
there is no essential difference: for any fixed value of $\hat\Phi_R$, the 
isovoltage $T_R-S$ curve contains two branches, one with positive isovoltage heat 
capacity and the other with negative isovoltage heat capacity. Therefore, 
there can be no first order isovoltage 
phase equilibrium. Once again, let us remark that the same qualitative behavior 
also appears in the asymptotic RPS formalism for RN-AdS BHs.
}

%%%%%%%%%%%%%%%%%%%%%%%%%%%%%%%%%%
\section{Concluding remarks}
\label{sec:Conclusion}
%%%%%%%%%%%%%%%%%%%%%%%%%%%%%%%%%

The work presented in this paper has two major achievements. 
Firstly, we extended the previous asymptotic RPS formalism for BHT into 
the quasi-local regime and illustrated in two example cases of RN and RN-AdS BHs 
that, for static observers located at a finite radial distances, 
the quasi-local RPS formalism possesses all necessary features of ordinary 
thermodynamic systems of which the Euler homogeneity is the key structure. 
Secondly, we analyzed the thermodynamic behaviors of the RN BHs in
the quasi-local RPS formalism and found that they behave extremely similar to the 
RN-AdS BHs in the asymptotic RPS formalism \cite{Zeyuan:2021uol}, but  
are significantly different to themselves in the asymptotic RPS formalism. 
The different behaviors of the RN BHs in the asymptotic and quasi-local formalisms
present an extra explicit example case for the common consensus that, 
in relativistic physics, different observers can observe different phenomena 
for the same underlying physical process.  

It is important to stress that the study of quasi-local formalism 
is still in its initial stage. Many related problems are awaiting for 
further explorations. First of all, what happens if the observer is at finite 
distance but is non-static? As is well known, in relativistic physics, 
the relation between thermodynamic quantities perceived by moving observers and 
those by static observers have been under debates for more than a hundred years,
see {\em e.g.} \cite{s41598-017-17526-4} for a brief history for the debates 
about the temperature perceived by moving observers. Using relativistic kinetic 
theory and the local version of Euler homogeneity, the transformation 
rule for the complete set of thermodynamic parameters for a relativistic gas 
is worked out in Ref.\cite{HLZ}. This may provide some clues for 
the quasi-local RPS formalism for spherically symmetric BHs 
from the view points of moving observers with purely radial relative velocities. 
However, the situation may become more complicated 
for moving observers with angular velocities, because the angular motion 
of the observers will break the spherical symmetry of the entire system 
including both the BHs and the observers.
Secondly, even for static observers, the quasi-local RPS formalism for 
non-spherically symmetric BHs is also an open issue. The central problem 
lies in how the world lines of the observers should be arranged in order to 
perceive a uniform temperature from all angular directions. 
Thirdly, the present work considers only spherically symmetric BHs in 
Einstein-Maxwell theory in four spacetime dimensions.
\blue{Other spherically symmetric BHs in the Brown-York quasi-local description
have been investigated with the absence of Euler relation\cite{Akbar:2004ke,Lee:2018hrd}.
Since the analysis in Sec.~\ref{sec:quasi} is not limited to RN black holes,
the extension to the
cases of other spherically symmetric BHs in other gravity models is 
also worth of consideration.}
The list of related open problems can be 
further prolonged, but we think the above is sufficient to 
illustrate the necessity of further explorations in this direction.

Notes Added: After this manuscript has appeared on arXiv, we were informed about the
interesting work \cite{Borsboom}, which also considered the quasi-local description of 
asymptotically AdS and flat BHs and the Euler homogeneity, however in the 
holographic extended phase space formalism proposed in \cite{visser2022holographic}. 
There are significant differences between their approach and ours. The complete
Euler homogeneity and the full Euler relation do not hold in the quasi-local 
description of the holographic extended phase space formalism, 
and the authors of \cite{Borsboom} termed the corresponding
variants as quasi-homogeneity and generalized Euler relation. Consequently, 
a key feature of extensive thermodynamic systems described by Eq.(3.15) is 
missing and the extensivity is broken
for asymptotically flat BHs in their approach. For comparison, 
the quasi-local RPS formalism
presented in this work is able to avoid all these issues.

%%%%%%%%%%%%%%%%%%%%%%%%%%%%%%%
\section*{Acknowledgement}
%%%%%%%%%%%%%%%%%%%%%%%%%%%%%%%
This work is supported by the National Natural Science Foundation 
of China under grant No. 12275138.

%%%%%%%%%%%%%%%%%%%%%%%%%%%%%%%
\section*{Data Availability Statement} 
%%%%%%%%%%%%%%%%%%%%%%%%%%%%%%%
This research makes no use of new data. 

%%%%%%%%%%%%%%%%%%%%%%%%%%%%%%%
\section*{Declaration of competing interest}
%%%%%%%%%%%%%%%%%%%%%%%%%%%%%%%
The authors declare no competing interest.

%%%%%%%%%%%%%%%%%%%%%%%%%%%%%%%%%%%%%%%%%%

%%%%%%%%%%%%%%%%%%%%%%%%%%%%%%%%%%%%%%%%%

\providecommand{\href}[2]{#2}\begingroup\raggedright
\providecommand{\eprint}[2][]{\href{http://arxiv.org/abs/#2}{arXiv:#2}}
\endgroup

\end{document}